\documentclass[sigconf]{acmart}

\usepackage{booktabs} 
\usepackage{url}
\usepackage{spverbatim}
\usepackage{endnotes}
\usepackage{hyperref}
\hypersetup{
    colorlinks = true,
    linkcolor = {blue},
    urlcolor = {blue},
    citecolor = {blue},
    hyperfootnotes={false}.
}

\usepackage{courier}

\usepackage{graphicx}
\graphicspath{ {images/} }

\usepackage{array}

\setcopyright{none}

\begin{document}
\title{Privacy Loss in Apple's Implementation of Differential Privacy on MacOS 10.12}

\author{Jun Tang}
\affiliation{%
  \institution{University of Southern California}
}
\email{juntang@usc.edu}

\author{Aleksandra Korolova}
\affiliation{%
  \institution{University of Southern California}
}
\email{korolova@usc.edu}

\author{Xiaolong Bai}
\affiliation{%
  \institution{Tsinghua University}
}
\email{bxl12@mails.tsinghua.edu.cn}

\author{Xueqiang Wang}
\affiliation{%
  \institution{Indiana University}
}
\email{xw48@indiana.edu }

\author{Xiaofeng Wang}
\affiliation{%
  \institution{Indiana University}
}
\email{xw7@indiana.edu}


\begin{abstract}
In June 2016, Apple made a bold announcement that it will deploy local differential privacy for some of their user data collection in order to ensure privacy of user data, even from Apple~\cite{applekeynote2016, apple709}. The details of Apple's approach remained sparse. Although several patents~\cite{thakurta2017emoji, thakurta2017learning, thakurta2017learning2} have since appeared hinting at the algorithms that may be used to achieve differential privacy, they did not include a precise explanation of the approach taken to privacy parameter choice. Such choice and the overall approach to privacy budget use and management are key questions for understanding the privacy protections provided by any deployment of differential privacy.

In this work, through a combination of experiments, static and dynamic code analysis of macOS Sierra (Version 10.12) implementation, we shed light on the choices Apple made for privacy budget management. We discover and describe Apple's set-up for differentially private data processing, including the overall data pipeline, the parameters used for differentially private perturbation of each piece of data, and the frequency with which such data is sent to Apple's servers.

We find that although Apple's deployment ensures that the (differential) privacy loss per each datum submitted to its servers is $1$ or $2$, the overall privacy loss permitted by the system is significantly higher, as high as $16$ per day for the four initially announced applications of Emojis, New words, Deeplinks and Lookup Hints~\cite{apple709}. Furthermore, Apple renews the privacy budget available every day, which leads to a possible privacy loss of 16 times the number of days since user opt-in to differentially private data collection for those four applications.

We applaud Apple's deployment of differential privacy for its bold demonstration of feasibility of innovation while guaranteeing rigorous privacy. However, we argue that in order to claim the full benefits of differentially private data collection, Apple must give full transparency of its implementation and privacy loss choices, enable user choice in areas related to privacy loss, and set meaningful defaults on the daily and device lifetime privacy loss permitted. 
\end{abstract}

%
%

\maketitle

\section{Introduction}\label{sec:intro}
Differential privacy~\cite{dwork2006calibrating} has been widely recognized as the leading statistical data privacy definition by the academic community~\cite{dwork2011acm, godel}. 
Thus, as one of the first large-scale commercial deployments of differential privacy (preceded only by Google's RAPPOR~\cite{rappor}), Apple's deployment is of significant interest to privacy theoreticians and practitioners alike. Furthermore, since Apple may be perceived as competing on privacy with other consumer companies, understanding the actual privacy protections afforded by the deployment of differential privacy in its desktop and mobile OSes may be of interest to consumers and consumer advocate groups~\cite{eff}.

However, Apple's publicly-facing communications about its deployment of differential privacy have been extremely limited: neither its developer documents~\cite{applenews, applenewsicloud, apple709, apple102, apple702} nor interstitials prompting the users to opt-in to differentially private data collection (Figures~\ref{fig:opt_in} and~\ref{fig:analytics_privacy}) provide details of the technology, except to say what data types it may be applied to. Two aspects of the deployment are crucial to understanding its privacy merits: the algorithms or processes used to ensure differential privacy of the data being sent and the privacy parameters being used by those algorithms. Although one can speculate about the algorithms deployed based on the recent patents~\cite{thakurta2017emoji, thakurta2017learning, thakurta2017learning2}, the question of parameters used to govern permitted privacy loss remains open and is our primary focus.

Both EFF and academics have called for Apple to detail its privacy budget use~\cite{eff, epsilonregistry1, barbosa, thakurtausenix}, to no avail\footnote{Apple's only public comments on the privacy budget are ``Restrict the number of submissions made during a period. No identifiers. Periodically delete donations from server"~\cite{apple709}.}. 
As far as we are aware, we are the first to systematically study privacy budget use in Apple's deployment of differential privacy.

\subsection{The (Differential) Privacy Budget}
One of the core distinctions of differential privacy (DP) from colloquial notions of privacy is that the definition provides a way to quantify the privacy risk incurred whenever a differentially private algorithm is deployed. Typically called \textit{privacy budget} or \textit{privacy loss} and denoted by $\epsilon$, it quantitatively measures by how much the risk to an individual privacy may increase due to that individual's data inclusion in the inputs to the algorithm. The higher the value of $\epsilon$, the less privacy protection is provided by the algorithm; in particular, the increase in privacy risks is proportional to $\exp(\epsilon)$. Although the choice of $\epsilon$ is typically treated as a social choice by the theoretical computer scientists~\cite{dwork2011acm}, it is of crucial importance in practical deployments, as the meaning of a privacy risk of $\exp(1)$ vs $\exp(50)$ is radically different.

In practice, an individual's data contribution is rarely limited to one datum. Whenever multiple data are submitted with differential privacy, the overall differential privacy loss incurred by that individual is viewed as bounded by the sum of the privacy losses of each of the submissions, due to what is known as composition theorems~\cite{dwork2014algorithmic, nontechnicaldp}. Hence, understanding the privacy implications of a deployed system such as Apple's, requires not only understanding the privacy loss incurred per datum submitted, but also how many datums may be submitted per time period or over a lifetime of a user's device. In fact, the need to understand the total privacy loss of differential privacy deployments has prompted Dwork and Mulligan to propose an ``Epsilon Registry"~\cite{epsilonregistry1}.

\subsection{Our Findings}
We find that although the privacy loss per datum is strictly limited to privacy budgets typically used in the literature, the daily privacy loss permitted by the implementation exceeds values typically considered acceptable by the theoretical community~\cite{hsu2014differential}, and the overall privacy loss per device may be unbounded (Section~\ref{sec:privacy-loss-findings}).

\section{Overview}\label{sec:overview}
\subsection{System Components}
We start by listing the components of the DP system on Mac OS we have identified:

\begin{itemize}
\item The differential privacy framework, located at \textsf{/System/\\Library/PrivateFrameworks/DifferentialPrivacy.framework}. \\The framework contains code implementing differential privacy, which we will decompile with Hopper Disassembler. In particular, it contains code responsible for per-datum privatization and for periodic functions that manage the privacy budget, updates of the database for privatized data, and creation of report files to be submitted to Apple servers.
\item The \textsf{com.apple.dprivacyd} daemon handling differential privacy, located at \textsf{/usr/libexec/dprivacyd}. We will study it using code tracing with LLDB.
\item A database, located at \textsf{/private/var/db/DifferentialPrivacy}, which contains several tables of privatized records and a table related to available budget per record type. Anyone with \textsf{sudo} privileges can open the database using \textsf{sqlite3}. We will study its contents (Section~\ref{sec:database}) and the changes to them due to usage of features that are supposed to trigger differentially private data collection and over time.
\item Configuration files (Figures~\ref{fig:keyname_plist}, \ref{fig:kp_plist}, \ref{fig:ap_plist} and \ref{fig:bp_plist}) with extension .plist, located at \textsf{/System/Library/DifferentialPrivacy/Configuration/}. The four files, which can be inspected by anyone but are difficult to change, specify numerous parameters that configure the actions of the DP framework, such as the per datum privacy parameter, privacy budget increase rate, etc. We will study the effects of each of these parameters by changing them and observing their effects during code execution with LLDB and through the resulting report files produced.
\item Report files (Figure~\ref{fig:report_file}) with extensions \textsf{.dpsub} and \textsf{.json.anon}, located at \textsf{/Library/Logs/DiagnosticReports/} and \\\textsf{/private/var/db/DifferentialPrivacy/Reports/}. These files contain privatized data and are the ones transmitted to Apple's servers. They can be opened with a text editor, and the \textsf{.dpsub} files can also be inspected through the MacOS Console under System Reports. We will study when they get created, their contents, and when they get deleted through observations and experiments.
\item The MacOS Console (Figure~\ref{fig:new_console}), which contains messages mentioning differential privacy, either in the library or process name. The messages are timestamped and easily readable, and are thus useful in noting certain system actions.
\end{itemize}

\subsection{System Organization and Data Pipeline}
The \textsf{dprivacy} (com.apple.dprivacyd) daemon runs the system responsible for implementation of differential privacy. Once a user opts-in to differentially private data collection in the MacOS Security \& Privacy Settings (Figure~\ref{fig:opt_in}), the \textsf{dprivacy} daemon is enabled 
and the database that will be supporting relevant data storage and management is created in \textsf{/var/db/DifferentialPrivacy}. Furthermore, there's a message visible on Console: ``dprivacyd: accepting work now". 

Per Apple's original announcement~\cite{applekeynote2016, apple709, applenews}, the use of DP is focused on four applications: new words, emojis, deeplinks, and lookup hints in Notes, with iCloud data added as an additional application in early 2017~\cite{applenewsicloud}, and further types of data collection such as health data introduced in mid-2017~\cite{apple702}. We observed how to reliably trigger DP-related activity when entering new words and emojis\footnote{To reliably trigger DP application to emojis, the user needs to call out the emoji keyboard by pressing "ctrl-cmd space", then click an emoji (or select an emoji with the arrow key and press "Enter"). For new words, the user can type an incorrectly spelled word in the Notes app and then ignore the spelling suggestion by pressing 'esc'.}; thus, our conclusions will be based on experiments with those applications. 

Whenever a user enters an emoji or a previously unseen new word in Notes, the relevant datum is perturbed using a differentially private algorithm and its privatized version and some metadata are added to a corresponding database table. 

A ReportGenerator task (Figure~\ref{fig:console}) is run periodically, at which point some records from the database are selected and written to report files (Figure~\ref{fig:report_file}), which are then transmitted to Apple's servers. The table rows corresponding to the selected records are ``marked as submitted" and eventually deleted from the database by a task.

There are several other periodic maintenance tasks, whose effects are: to delete records from the database (even those that weren't submitted) and to delete report files from disk. These periodic tasks are accompanied by messages observable on the Console (Figure~\ref{fig:new_console}).

\subsection{Study Questions}\label{sec:questions}

In order to understand the privacy loss in Apple's implementation of differential privacy, we need to understand the following aspects of the system:
\begin{enumerate}
\item What are the privacy parameters used in order to achieve privatization before the privatized datum gets entered into the database? This will let us understand per datum privacy.
\item How frequently are records selected for inclusion in a report? How many records can be included in one report? How frequently are the reports created and submitted? This will let us understand the rate of privacy loss.
\item Is the total privacy loss that a particular user can incur limited?
\item How easy is it to alter the performance of the system, e.g., change parameters responsible for each datum's privatization, change the number of records selected for inclusion into a report or the frequency of report generation? 
\end{enumerate}

We discovered that the answers to questions (1) -- (3) (see Section~\ref{sec:privacy-loss-findings}) depend on the parameters specified in the configuration files and their use by the framework to establish the available privacy budget. We describe our findings and observations regarding the database tables, configuration files, and the functionality related to report generation and privacy budget maintenance next. We will discuss (4) in Section~\ref{sec:protections-epsilon}.

\section{System's Details}\label{sec:core}
\subsection{The Database}\label{sec:database}
\begin{figure}[h!]
\includegraphics[scale=0.4]{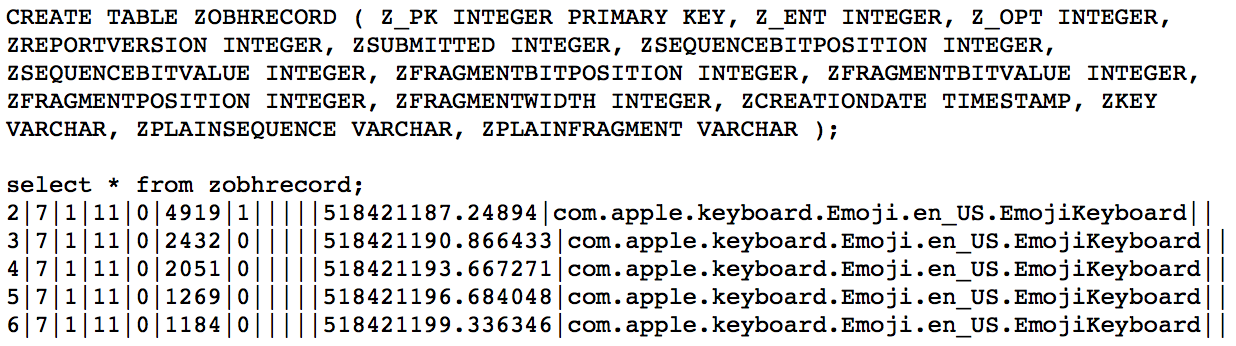}
\centering
\caption{Schema and entries in the ZOBHRECORD table.}
\label{fig:zobh_tb}
\vspace{-3em}
\end{figure}

The ZOBHRECORD (See Figure~\ref{fig:zobh_tb}) and ZCMSRECORD tables in the database store the perturbed data, with the former dedicated to the privatized emoji records, and the latter -- to the privatized new words records. Every emoji typed by a user gets privatized and stored in the ZOBHRECORD table. In contrast, only words that haven't been previously typed are privatized and stored in the ZCMSRECORD table.

\begin{figure}[h!]
\includegraphics[scale=0.42]{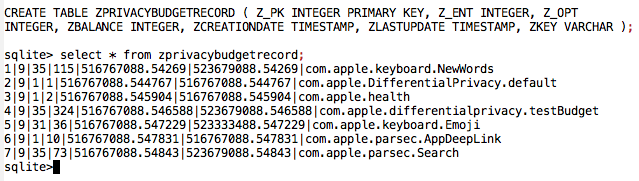}
\centering
\caption{Schema of and privacy budget items from the ZPRIVACYBUDGETRECORD table.}
\label{fig:zp_tb}
\end{figure}

A notable table is ZPRIVACYBUDGETRECORD, whose schema and example contents are shown in Figure~\ref{fig:zp_tb}. The table contains 7 entries, one for each of the applications (NewWords, Emoji, AppDeepLink, Search, and health) and two for helper functions (default and testBudget). The ZBALANCE column contains the integer value of the currently available privacy budget for each application.

\subsection{Configuration Files}
\begin{figure}[h]
\includegraphics[scale=0.4]{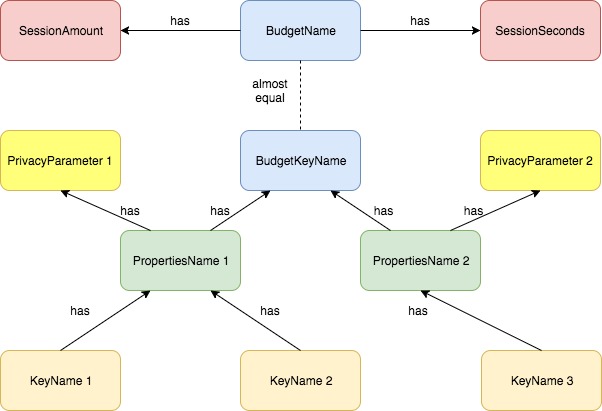}
\centering
\caption{Relation between KeyName, PropertiesName, BudgetKeyName, BudgetName, etc.}
\label{fig:names}
\end{figure}

\begin{figure}[h]
\includegraphics[scale=0.55]{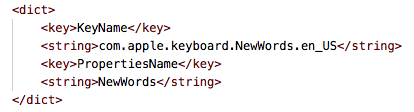}
\centering
\caption{Example of keynames.plist}
\label{fig:keyname_plist}
\end{figure}

\begin{figure}[h]
\includegraphics[scale=0.55]{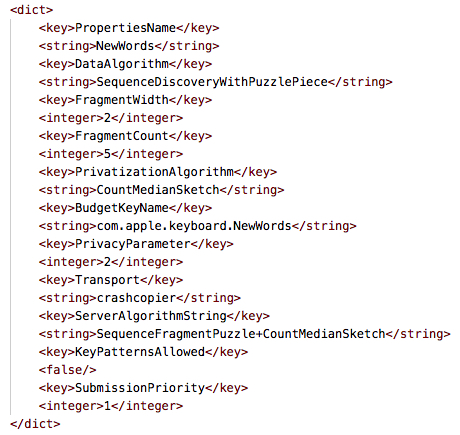}
\centering
\caption{Example of keyproperties.plist}
\label{fig:kp_plist}
\vspace{-2em}
\end{figure}

\begin{figure}[h]
\includegraphics[scale=0.55]{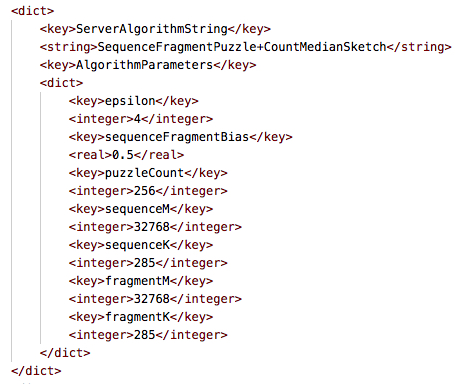}
\centering
\caption{Example of algorithmparameters.plist}
\label{fig:ap_plist}
\end{figure}

There are four configuration files for the DP daemon: com.apple.dprivacyd.\\\{\textit{keynames, keyproperties, algorithmparameters, budgetproperties}\}.plist. 
See Figures~\ref{fig:keyname_plist}, \ref{fig:kp_plist}, \ref{fig:ap_plist} and \ref{fig:bp_plist} for snippets of the configuration files, Figure~\ref{fig:names} for a schematic relationship between keys in them, and Tables~\ref{table:names} and~\ref{table:sessionamount} for a snippet of their values.

\subsubsection{KeyName $\rightarrow$ PropertiesName}
\textit{keynames.plist} contains a mapping of KeyNames to PropertiesNames.
KeyNames describe the possible data types, e.g., \\com.apple.keyboard.NewWords.en\_US -- a new word in English using the US keyboard\footnote{We are not certain whether the second identifier encodes a keyboard preference or a region preference, or both.}, com.apple.keyboard.NewWords.en\_GB -- a new word in English using the Great Britain keyboard, \\com.apple.keyboard.NewWords.ru\_RU -- a new word in Russian, com.apple.keyboard.Emoji.fr\_FR.EmojiKeyboard -- an emoji in French, com.apple.parsec.AppDeepLink -- a deeplink. In MacOS 10.12.3 \textit{keynames.plist} contains 160 distinct KeyNames.

Each KeyName is assigned one of 13 possible PropertiesName.
For example, KeyName com.apple.keyboard.NewWords.en\_US has a PropertiesName NewWords, as do com.apple.keyboard.NewWords.en\_GB and com.apple.keyboard.NewWords.ru\_RU; \\ KeyName com.apple.parsec.AppDeepLink has a PropertiesName DeepLinks; KeyName com.apple.keyboard.Emoji.fr\_FR.EmojiKeyboard has a PropertiesName TermFrequency, as do \\ com.apple.keyboard.Emoji.ru\_RU.EmojiKeyboard and \\com.apple.keyboard.Emoji.en\_US.EmojiKeyboard.

\subsubsection{Determining the Per-Datum Privacy Loss: KeyName $\rightarrow$ PropertiesName $\rightarrow$ PrivatizationAlgorithm, PrivacyParameter}
For each of the 13 possible PropertiesName values, the \textit{keyproperties.plist} file specifies a PrivatizationAlgorithm and PrivacyParameter. For example, for PropertiesName=HealthDataTypes: PrivatizationAlgorithm=OneBitHistogram and PrivacyParameter=1; for PropertiesName=NewWords: ~PrivatizationAlgorithm=~CountMedianSketch and PrivacyParameter=2; for ~PropertiesName=~TermFrequency: PrivatizationAlgorithm=OneBitHistogram and PrivacyParameter=1. 

\textit{algorithmparameters.plist} specifies additional parameters of the privatization algorithm.

\subsubsection{Determining a Budget for Particular Data Types: KeyName $\rightarrow$ PropertiesName $\rightarrow$ BudgetKeyName}
Furthermore, for each of the 13 possible PropertiesName values, the \textit{keyproperties.plist} file specifies a BudgetKeyName. 
For example, for PropertiesName=LocalWords: BudgetKeyName=com.apple.keyboard.NewWords; for PropertiesName=NewWords: BudgetKeyName=com.apple.keyboard.NewWords; for 
~PropertiesName=~DeepLinks: ~BudgetKeyName=~com.~apple.~parsec.~AppDeepLink; for ~PropertiesName=~TermFrequency: BudgetKeyName=com.apple.keyboard.Emoji. 
In particular, this implies that all new words, regardless of language, have the same ~BudgetKeyName=com.apple.~keyboard.~NewWords, all emojis -- the same BudgetKeyName=com.~apple.keyboard.Emoji,
 etc. There are a total of 7 distinct BudgetKeyNames, which is consistent with what we saw in the ZPRIVACYBUDGETRECORD table in the database (Figure~\ref{fig:zp_tb}).

\begin{table*}[h!]
\centering
\begin{tabular}{|p{5.1cm} | p{2.3cm}| p{2.5cm} | p{1.2cm} | l | p{1cm} |}
    \hline
      KeyName & PropertiesName & Privatization-Algorithm & Privacy-Parameter & BudgetName & Session-Amount \\
     \hline
     com.apple.keyboard.NewWords.it\_IT & NewWords & CountMedianSketch & 2 & com.apple.keyboard.NewWords & 2 \\
     \hline
    com.apple.keyboard.NewWords.ru\_RU & NewWords & CountMedianSketch & 2 & com.apple.keyboard.NewWords & 2 \\
     \hline
      com.apple.keyboard.NewWords.zh\_Hans & NewWordsChinese & CountMedianSketch & 2 & com.apple.keyboard.NewWords & 2 \\
    \hline
    com.apple.keyboard.LocalWords.en\_US & LocalWords & CountMedianSketch & 2 & com.apple.keyboard.NewWords & 2 \\
    \hline
    com.apple.keyboard.Emoji.fr\_FR.Emoji & TermFrequency & OneBitHistogram & 1 & com.apple.keyboard.Emoji & 1 \\
         \hline
    com.apple.parsec.AppDeepLink & DeepLinks & CountMedianSketch & 1 & com.apple.parsec.AppDeepLink & 10 \\
         \hline
    com.apple.health.datatypes & HealthDataTypes &OneBitHistogram & 1 & com.apple.health & 2 \\
         \hline
    com.apple.lookup.QueryMatch.~queryOnly.notHighlighted & Search & OneBitHistogram & 1 & com.apple.parsec.Search & 1 \\
         \hline
    com.apple.lookup.DomainMatch & Search & OneBitHistogram & 1 & com.apple.parsec.Search & 1 \\ 
\hline
\end{tabular}
\caption{Values for particular KeyNames: PropertiesName, PrivatizationAlgorithm, PrivacyParameter and BudgetName}
\label{table:names}
\end{table*}

\begin{table}[h!]
\centering
\begin{tabular}{|m{5cm} | p{1cm} | p{1cm} |}
\hline
BudgetKeyName & Session-Seconds & Session-Amount \\
\hline
com.apple.keyboard.Emoji & 86400 & 1 \\
com.apple.parsec.Search & 86400 & 1 \\
com.apple.keyboard.NewWords & 86400 & 2 \\ 
com.apple.parsec.AppDeepLink & 86400 & 10 \\
com.apple.health & 604800 & 2 \\
com.apple.differentialprivacy.testBudget & 86400 & 4 \\
com.apple.DifferentialPrivacy.default & 86400 & 1\\
\hline
\end{tabular}
\caption{Budget Properties as specified in \textit{budgetproperties.plist} configuration file on MacOS 10.12.3}
\label{table:sessionamount}
\end{table}

\begin{figure}[h]
\includegraphics[scale=0.55]{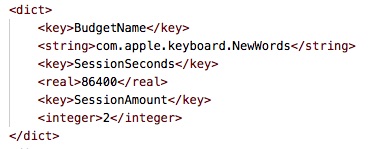}
\centering
\caption{Example of budgetproperties.plist}
\label{fig:bp_plist}
\end{figure}

\subsubsection{Budget Properties}
The \textit{budgetproperties.plist} file specifies two quantities for each BudgetKeyName: SessionSeconds and SessionAmount (see Figure~\ref{fig:bp_plist} for a snippet and Table~\ref{table:sessionamount} for the values).

SessionSeconds = 86400 for all BudgetKeyNames except \\ com.apple.health, for whom SessionSeconds = 604800. These correspond to number of seconds in a day and in a week, respectively.

SessionAmount values range from 1 to 10, depending on BudgetKeyName (see Table~\ref{table:sessionamount}).

\section{Privacy Loss Findings}\label{sec:privacy-loss-findings}
We now answer the questions posed in Section~\ref{sec:questions}.

\subsection{Each Datum's Privatization}
\textbf{PrivacyParameter of the KeyName specifies the privacy parameter used for privatization of the datum of that KeyName prior to its addition to the database.}

For example, for an emoji in French, Russian, or English, the privatization algorithm that will be run is OneBitHistogram with PrivacyParameter 1. For a new word in English using the US keyboard or a new word in Russian, the privatization algorithm that will be run is CountMedianSketch with PrivacyParameter 2, etc. See Table~\ref{table:names} for the PrivacyParameter values used for the various datum types in Mac OS 10.12.3.

It is difficult to understand and verify correctness of the privatization algorithms when one only has access to the binary code. We expect that the algorithms implement the ideas described in the patents~\cite{thakurta2017emoji, thakurta2017learning, thakurta2017learning2}.
What we do verify using LLDB and by changing the PrivacyParameter in the configuration file and observing the effects, is that the privacy parameter \textsf{epsilon} used for emoji and new words privatization is what one would expect based on the values in the configuration file (see Section~\ref{sec:app-1}).

\subsection{Report Generation and Privacy Budget Management over Time}\label{sec:rgpbm}
\textbf{SessionAmount specifies how many records belonging to a particular KeyName can be included in a report file. SessionAmount also specifies the increase to the available budget balance for each BudgetKeyName that happens every SessionSeconds.}

\subsubsection{Number of Records per Report}
Specifically, Apple keeps track of the privacy budget balance (ZBALANCE) available for each of the 7 BudgetKeyNames in the ZPRIVACYBUDGETRECORD database table. The available budget balance together with the SessionAmount is used to decide how many records to include in each report file. Specifically, every 18 hours, the daemon runs a ReportGenerator task. It selects records from the database tables to be included in the report according to the following:
\begin{itemize}
\item At most $\min$(SessionAmount, 40)\footnote{The number 40 is hard-coded in the binary code (Figure~\ref{fig:min40}).} records per KeyName may be selected.
\item The total number of records belonging to the same BudgetKeyName selected may not exceed the privacy budget balance for that BudgetKeyName currently available as per the ZPRIVACYBUDGETRECORD table in the database. When the number of records of a particular BudgetKeyName available in the database tables exceeds the available budget balance, then the subset of records whose total number does not exceed the available budget balance are chosen at random while taking into account SubmissionPriority.  
\end{itemize}

Each record selected is ``marked as submitted" in the corresponding table in the database, and for each record submitted the corresponding privacy budget balance is decreased by one.

The report files created contain the creation time (or creation time adjusted forward by 7 hours) in the file name, and are placed in the folder \textsf{/Library/Logs/DiagnosticReports/} or \textsf{/private/var/db/~DifferentialPrivacy/Reports/}. Records with KeyNames that have TermFrequency, NewWords or LocalWords as their PropertiesName are included in reports in the first folder;  records with KeyNames that have Search as PropertiesName -- in the second folder.

\subsubsection{Budget Increase}
A periodic task PrivacyBudgetMaintenance increases the ZBALANCE value in the ZPRIVACYBUDGETRECORD table for each BudgetKeyName by its corresponding SessionAmount every SessionSeconds (see Section~\ref{sec:SessionAmountSignificance} for experimental and code evidence supporting this claim). Thus, for all BudgetKeyNames except health and default, the available privacy budget balance is increased by SessionAmount every 24 hours\footnote{The default and AppDeepLink BudgetKeyNames are the exceptions to this; the former likely due to its role as a default value for abuse scenarios (Section~\ref{sec:protections-epsilon}) and the latter due to DeepLink functionality not being present in MacOS 10.12.}. Due to OS sleep, which is common on MacOS, in practice the daemon increases the privacy budget balance by the SessionAmount multiplied by the number of days that have elapsed since the last budget update (a time which is kept track of in the database table).

When a user opts-in to differentially private data collection, the budgets for each BudgetKeyName are initialized with their corresponding SessionAmount.

\subsubsection{Total Privacy Loss Permitted}
Consider an example: a user opts in to DP, then types one or several emoji every day for $t$ days. Since for emoji PrivacyParameter=1, every emoji will be put into database after being privatized with privacy loss of 1. Since for emoji the SessionAmount=1, the privacy budget balance will be increased by one every day, and so every day one privatized emoji will be included in a report sent to Apple's servers. After $t$ days, by composition theorems~\cite{dwork2014algorithmic, nontechnicaldp}, the privacy loss incurred will be $1 \cdot 1 \cdot t = t$.

Consider another example: a user opts in to DP, but doesn't use any emoji for 20 days. The available privacy budget balance for emoji will be 20 after that time. Then the user types two emoji each in 10 different languages supported by differential privacy in one day. Each of the 20 emojis will be put into the database after being privatized with privacy loss of 1. In the first day, 10 different emoji, one from each language (since SessionAmount = 1 for each emoji KeyName and available privacy budget balance is 20), will be included in the report, for a privacy loss of 10. The next day, the remaining 10 different emoji will be included in the report, for an additional privacy loss of 10.

In other words, the privacy loss for a particular application permitted by the implementation is PrivacyParameter $\cdot$ SessionAmount every SessionSeconds. The privacy loss that is not realized during particular time period if that application is not used remains available for future use.
Thus, for the applications of NewWords, AppDeepLink, Search, and Emoji, whose respective PrivacyParameters are: 2, 1, 1, 1,  and SessionAmounts are: 2, 10, 1, 1, and SessionSeconds is 86,400 in MacOS 10.12.3, the overall daily privacy loss permitted is 16. Moreover, since unused privacy budget balance rolls over for subsequent use, the overall privacy loss of a device for the four initially announced applications by Apple may reach 16 times the number of days since the user of the device has opted in to DP. A caveat to these findings is that DeepLink functionality does not appear to be implemented on MacOS yet, so the actual privacy loss on Mac OS 10.12.3 is currently as large as 6 per day and on iOS 10.1.1 -- as large as 14 per day (Section~\ref{sec:ios}) for the four initially announced applications.
\section{Discussion}\label{sec:discussion}

\subsection{Report File and Database Maintenance}
Besides the periodic tasks of ReportGenerator and PrivacyBudgetMaintenance, whose actions have already been described in Section~\ref{sec:rgpbm}, the following 3 periodic tasks are responsible for database and report file management (Figure~\ref{fig:console}):
\begin{itemize}
\item StorageCulling (every 24 hours): deletes records that have been submitted and records with the mismatched version number from the database.
\item StorageMaintenance (every 12 hours): deletes records to limit database size and deletes records added to the database more than two weeks before the current date.
\item ReportFilesMaintenance (every 24 hours): removes report files older than a month\footnote{Concluded based on observation and dynamic code analysis, as we could not find this in the framework code.} from disk.
\end{itemize}

\subsection{Ease of Altering System's Performance}\label{sec:protections-epsilon}
We have observed several precautions that are implemented by Apple in order to make it difficult to abuse the implementation: 
\begin{itemize}
\item The configuration files 
are difficult to change, as such a change on Mac OS requires turning off Apple's System Integrity Protection, which is not trivial. We have not found a way to change the configuration files on iOS.
\item Even if one succeeds in changing the configuration files, whenever a PrivacyParameter in a configuration file is set to a value higher than epsilonMax, a constant value equal to 2 which is hard-coded in the code of the framework (Figure~\ref{fig:hardcoded}), the PrivacyParameter used is defaulted to 1 at runtime  and the corresponding record's Submission Priority is set to 99999, effectively ensuring it does not get included in report files. Furthermore, SessionAmount is defaulted to at most 40 at runtime (Section~\ref{sec:rgpbm}).
\item Time measures used by the daemon, such as the number of seconds in 18 hours, in a day, in 7 days, are hardcoded in the code (Section~\ref{sec:hardcoded}). That may be the reason why we have not succeeded in accelerating report file generation or privacy budget increase by changing the SessionSeconds in the configuration file or changing the computer's clock.
\end{itemize}
On the other hand, anyone with root permissions can alter the privacy budget balance in the database, thereby artificially increasing the privacy loss.

\subsection{Configuration Differences between MacOS versions}
We observed that Apple made changed to configuration files from MacOS 10.12.1 to 10.12.3 (see Table~\ref{table:diff}). The main distinctions are that in MacOS 10.12.3:
\begin{itemize}
\item \textsf{SessionAmount} for \textsf{com.apple.keyboard.NewWords} increases from 1 to 2, resulting in a higher daily privacy loss.
\item \textsf{BudgetName} \textsf{com.apple.health} and \textsf{PropertiesName} \textsf{LocalWords} are introduced, signaling new applications for DP.
\item SubmissionPriority is adopted, signaling new protections against abuse.
\item Health-related PropertiesName and Budget are introduced, signaling that health-related data will also be included in DP data collection.
\end{itemize}

\begin{table}[h]
\centering
\begin{tabular}{| m{5.8cm} | m{0.75cm}| m{0.75cm} |}
    \hline
     & MacOS 10.12.1 &  MacOS 10.12.3 \\
     \hline
     SessionAmount for NewWords & 1 & 2 \\
     \hline
     SessionAmount for testBudget & 1 & 4 \\
     \hline
     BudgetName com.apple.health & no & yes \\
    \hline  
    PropertiesName LocalWords & no & yes \\
     \hline
    apple.photos.search.miss.unnormalized.~en\_US & yes & no \\
    \hline
    apple.photos.search.miss.normalized.en\_US & yes & no  \\
    \hline
    PropertiesName HealthDataTypes & no & yes \\
    \hline
    PropertiesName LocalWords & no & yes \\
    \hline
    SubmissionPriority & no & yes \\
    \hline
\end{tabular}
\caption{Configuration file difference between MacOS 10.12.1 and 10.12.3}
\vspace{-2em}
\label{table:diff}
\end{table}

\subsection{A Note on iOS}\label{sec:ios}
Studying the iOS DP implementation is significantly more difficult. From our observations of the Console messages when the iPhone is connected to a computer running MacOS and of the reports (observable under Settings $\rightarrow$ Privacy $\rightarrow$  Analytics $\rightarrow$  Analytics Data), the iOS implementation follows the same principles as the MacOS one. The configuration files for iOS 10.1.1 we obtained from a jailbroken phone were identical to those of MacOS 10.12.1. The distinctions we found relate to iOS reports containing more metrics (in particular, we were not able to trigger DeepLink functionality on MacOS while such records abound on iOS), and to faster report file deletion from the phone than from the computer (7 vs 30 days).

\section{Conclusions and Future Work}\label{sec:conclusions}
We applaud Apple for its deployment of differential privacy in the local privacy model and for the many safeguards put in place to make it difficult to abuse. However, we believe the deployment has several significant shortcomings.
\begin{enumerate}
\item The privacy loss permitted by the system is not explained anywhere and takes significant effort to reverse-engineer. This is contrary to one of the main conceptual advantages of differential privacy -- that a user can make an informed choice whether to opt-in to differentially private data collection based on the quantifiable knowledge of risk announced by the data collector. 
\item Furthermore, the lack of transparency on privacy loss opens the door for intentional or un-intentional abuse by Apple itself, e.g., by unilaterally changing either the per-datum privacy loss or the rate of privacy loss in a time period or by introducing additional BudgetKeyName(s), Apple may significantly weaken the privacy guarantees provided without anyone's knowledge or consent. In fact, this may already be happening -- by inspecting iOS 11 beta report files (Figure~\ref{fig:report_fileiOS11}), we have observed that the daily privacy loss permitted increased by at least $29$ compared to that of iOS 10.1.1 and MacOS 10.12.3:
\begin{itemize}
\item the PrivacyParameter used for Emoji is $2$ (instead of $1$) and as many as $10$ records per Emoji KeyName (instead of $1$ as in iOS 10.1.1 and MacOS 10.12.3) are included in one report file;
\item additional KeyNames, such as com.apple.safari.DomainVisited and com.apple.safari.DomainCausingEnergyDrain, are introduced, each using a PrivacyParameter of $1$ and permitting as many as $10$ records for their BudgetKeyName per report file.
\end{itemize}
\item The privacy loss of 16 per day permitted by the system is significantly higher than what is commonly considered reasonable in academic literature. Furthermore, since the permitted privacy loss balance is replenished every day, over a course of time the total privacy loss per device becomes larger by orders of magnitude.
\item Due to the way the database and report files are structured, the implementation leaks what features of MacOS a user is using and in what language and, possibly, with what geographic and keyboard preference, both to Apple and to anyone who has access to the database. Furthermore, because only new words are privatized and added to the relevant database tables, one can potentially test whether a particular non-dictionary word has been ever used by the owner of the device by observing whether typing it triggers changes to the DP database.
\end{enumerate}
We call for Apple to make its implementation of privacy-preserving algorithms public and to make the rate of privacy loss fully transparent and tunable by the user.

\subsubsection{Future Work}
Apple does not transmit any user or device identifiers along with reports~\cite{apple709}. It would be worthwhile to investigate the effect that such (or other ways of) decoupling of data source from data aggregator can play in mitigating the implications of the (theoretically) infinitely increasing privacy loss~\cite{MironovPrivate}.

It has been observed that properly implementing algorithms claiming to preserve differential privacy is non-trivial in practice~\cite{MironovLeast, Barthe2016, Barthe2014}. It would be worthwhile to develop further techniques for verifying correctness of claimed DP implementations.

Finally, the question of how to intuitively interpret and convey the privacy guarantees and limitations of differential privacy at various privacy loss levels to the public, remains open. 

\bibliographystyle{ACM-Reference-Format}
\bibliography{apple-impl-bibliography} 

\appendix
\section{Appendix}\label{sec:appendix}
%
%
%
\subsection{Code Support for Findings}
The code convention for Objective-C, which Hopper disassembles the framework code to, is as follows. In an Objective-C function name, the leading \texttt{-} means this is an instance method that can 
only be accessed by an instance of the class, while \texttt{+} indicates the method is a class method and can be accessed anytime by simply referencing the class. The following 
 \texttt{[]} includes both the function name and argument name, and arguments are separated by \texttt{:} (an example can be found in Figure~\ref {fig:random_hopper}). 

\subsubsection{Checking PrivacyParameter corresponds to the privacy parameter used in a datum's privatization}\label{sec:app-1}
\begin{figure}[h]
\includegraphics[scale=0.5]{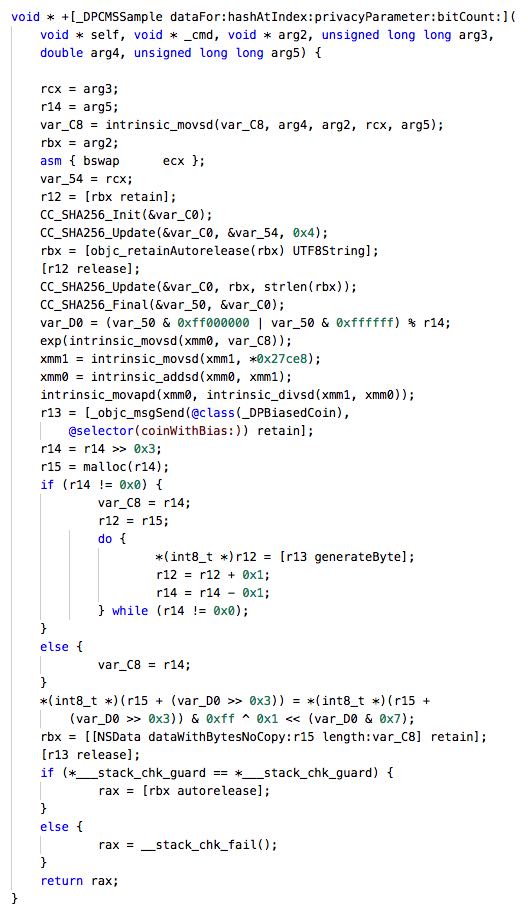}
\centering
\caption{Code snippet from Hopper for new word randomization (partial).}
\label{fig:cms_random_hopper}
\end{figure}

\begin{figure}[h]
\includegraphics[scale=0.6]{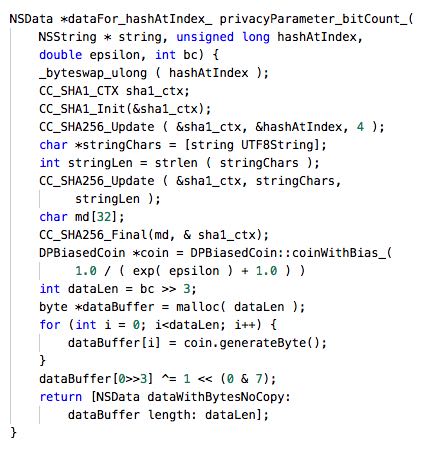}
\centering
\caption{Our interpretaion of the code snippet for new word randomization (partial).}
\label{fig:cms_random_an}
\end{figure}

\begin{figure*}[h]
\includegraphics[scale=0.23]{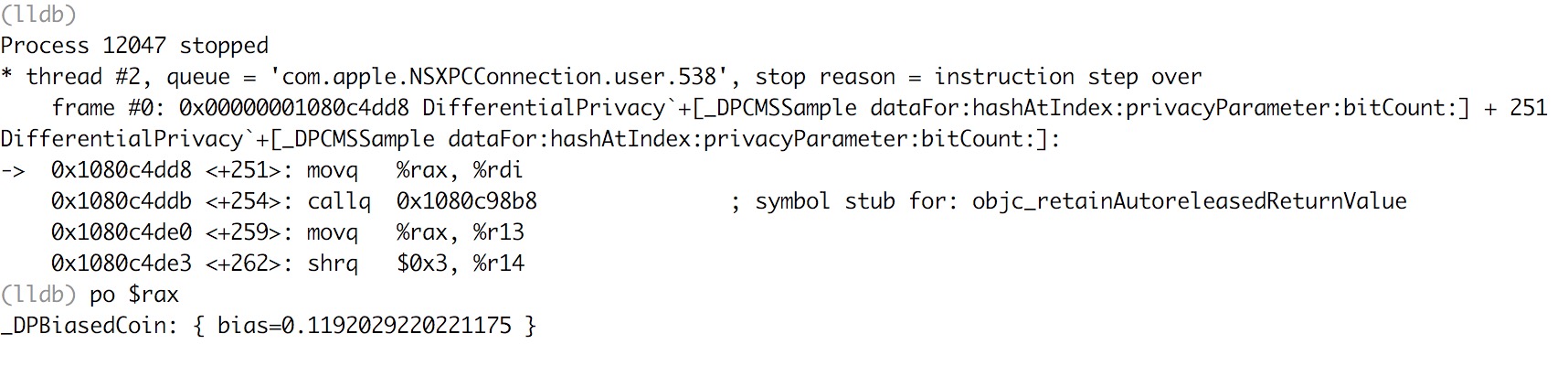}
\centering
\vspace{-2em}
\caption{\_DPBiasedCon value from LLDB.}
\label{fig:cms_random_lldb}
\end{figure*}

\begin{figure*}[h]
 \includegraphics[scale=0.45]{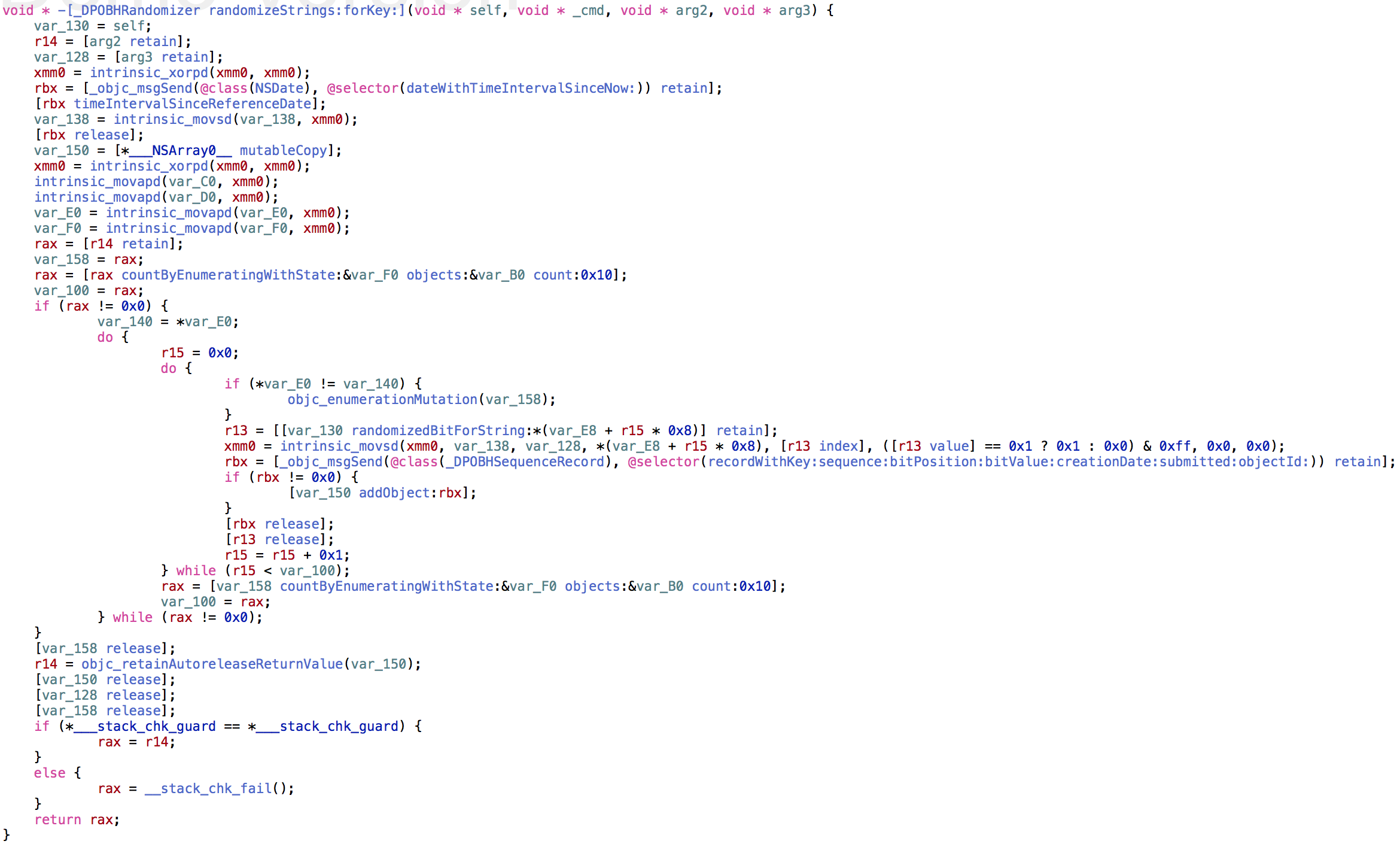}
 \centering
  \vspace{-3em}
 \caption{The emoji randomization code (partial) from Hopper Disassembler.}
 \label{fig:random_hopper}
 \vspace{-6em}
 \end{figure*}

For new words, we observed that the function \textsf{\_DPCMSSample initWith} uses PrivacyParameter to create the \textsf{\_DPBiasedCoin}
(see Figure~\ref{fig:cms_random_hopper} for Hopper code and Figure~\ref{fig:cms_random_an} for our interpretation of it), and using runtime LLDB we observed that the value is $1.0 / (\exp(PrivacyParameter) + 1.0)$ (Figure~\ref{fig:cms_random_lldb}).
Analogously, we analyzed the emoji randomization code (Figure~\ref{fig:random_hopper}). 

\subsubsection{Checking that SessionAmount controls the daily budget increase value}\label{sec:SessionAmountSignificance}
\begin{figure}[h]
\includegraphics[scale=0.6]{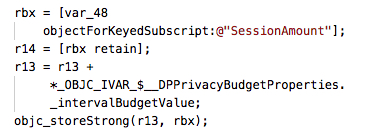}
\centering
\caption{Code evidence that connects SessionAmount to intervalBudgetValue.}
\label{fig:sessionamount_to_intervalbudgetval}
\end{figure}

\begin{figure}[h]
\includegraphics[scale=0.6]{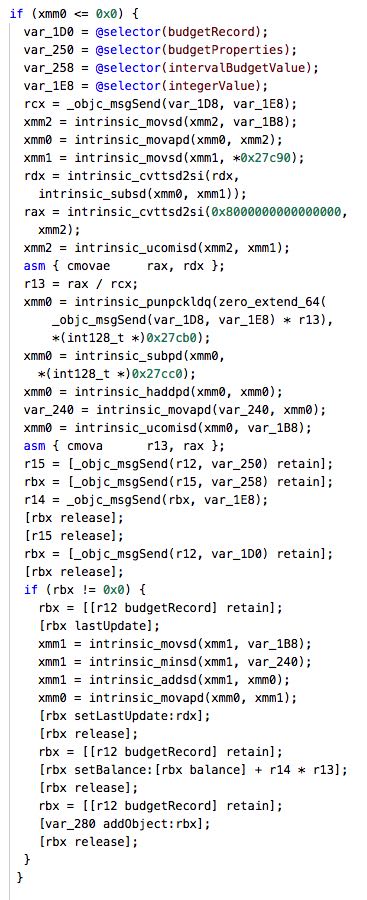}
\centering
\caption{Code evidence that connects intervalBudgetValue to daily budget increase value.}
\label{fig:r14_to_interbudgetval}
\end{figure}

In the \textsf{\texttt{-[}\_DPPrivacyBudgetProperties\ initWithDictionary\texttt{:]}} function (Figure~\ref{fig:sessionamount_to_intervalbudgetval}), the SessionAmount is assigned to intervalBudgetValue. 
In the \textsf{\_DPPrivacyBudget\ updateAllBudgetsIn} function (Figure~\ref{fig:r14_to_interbudgetval}), we can see the value used to multiply with the number of days since ZLASTUPDATE is r14,
and \\
 \texttt{r14 = (r15 intervalBudgetValue) interValue)}. \\
 So we can conclude that SessionAmount is used as the daily budget increase value.

\subsubsection{Hard-coded values}\label{sec:hardcoded}
\begin{figure}[h]
\includegraphics[scale=0.75]{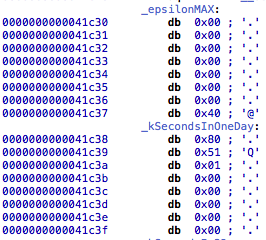}
\centering
\caption{epsilonMAX (equals to 2 if interpreted as a double type number) and number of seconds in one day are hardcoded in the code.}
\label{fig:hardcoded}
\end{figure}

\begin{figure}[h]
\includegraphics[scale=0.4]{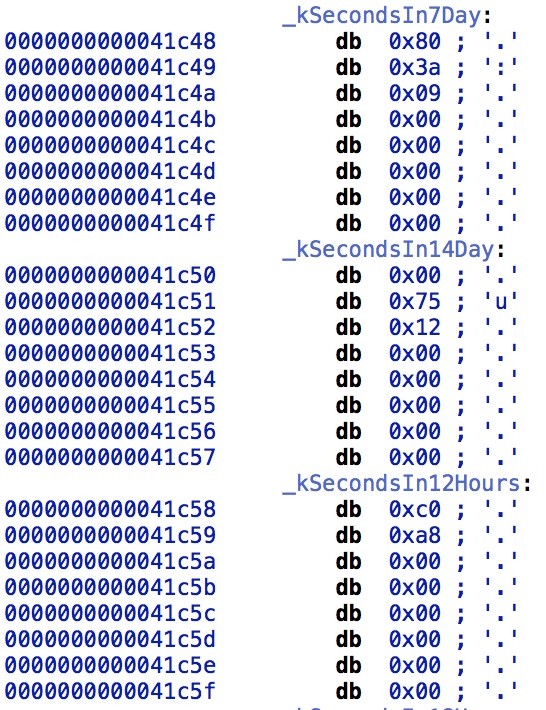}
\centering
\caption{Example of other hardcoded variables.}
\label{fig:other_hardcoded}
\end{figure}

In addition to epsilonMAX and \_kSecondsInOneDay (Figure~\ref{fig:hardcoded}), the following constants are also hardoced in the framework code: \_kSecondsIn3Day,  \_kSecondsIn7Day,  \_kSecondsIn14Day,  \_kSecondsIn12Hours,
 \_kSecondsIn18Hours, \_kSecondsIn24Hours (Figure~\ref{fig:other_hardcoded}).


%
%
%
%
\subsection{Aspects of the System that we don't Understand}
\begin{figure}[h]
\includegraphics[scale=0.35]{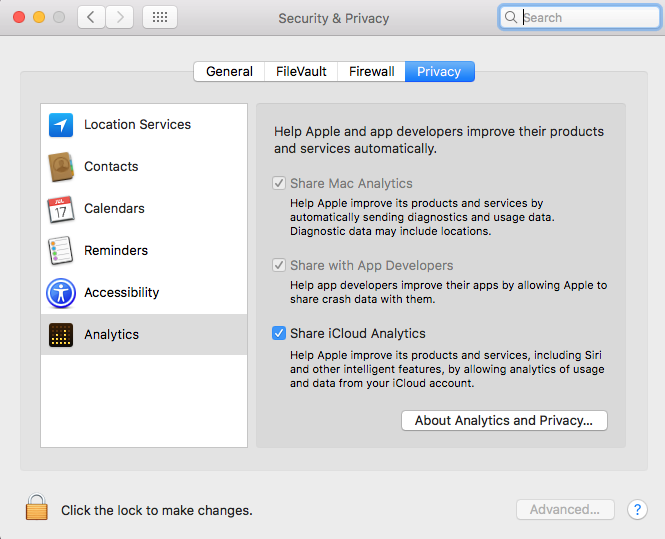}
\centering
\caption{Screenshot of the opt-in interface.}
\label{fig:opt_in}
\end{figure}

Several details of the implementation's behavior were puzzling:
\begin{itemize}
\item The privacy budget balance sometimes changes dramatically. We observed this phenomenon twice over the course of 6 months. In both cases, the privacy budget increased and was set to $ \frac{ZLASTUPDATE - ZCREATIONDATE} {86400}$.
We don't know the reason for this; one possibility is that the Apple server triggered this change remotely.
\item The daemon occasionally automatically stops data collection and un-checks the boxes under Settings $\rightarrow$ Security \& Privacy (Figure~\ref{fig:opt_in}) indicating opt-in to Analytics Sharing (and hence, differential privacy). Over the course of our experiments in the last 6 months, we observed this effect several times. We were not able to reliably reproduce it. In particular, our experiments of setting the privacy budget balance to zero or a negative number in the database or entering thousands of emoji within a short period of time did not trigger opt-out.
\item When PrivacyParameter $>$ epsilonMAX (Section~\ref{sec:protections-epsilon}), the NewWords record will be inserted into the ZOBHRECORD table, even though it is typically inserted into the ZCMSRECORD table.
\item We don't understand the role of testBudget, one of the 7 BudgetKeyNames. 
\end{itemize}

\subsection{Figures}

\begin{figure}[h]
\includegraphics[scale=0.35]{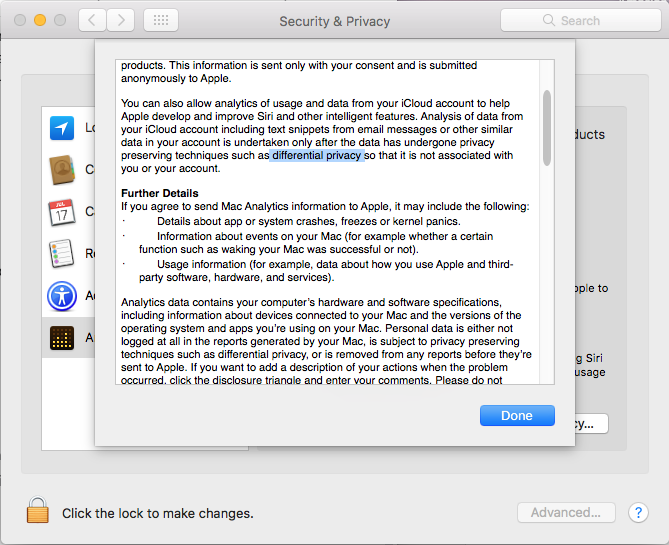}
\centering
\caption{Screenshot of Apple mentioning DP in ``About Analytics and Privacy".}
\label{fig:analytics_privacy}
\end{figure}

\begin{figure}[h]
\includegraphics[scale=0.48]{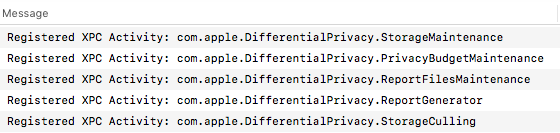}
\centering
\caption{Five periodic maintenance tasks.}
\label{fig:console}
\end{figure}

\begin{figure}[h]
\includegraphics[scale=0.6]{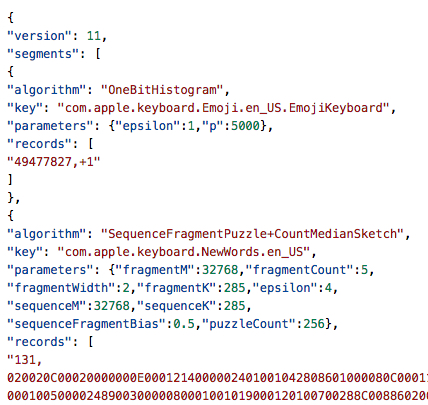}
\centering
\caption{An example report file (located in \textsf{/Library/Logs/DiagnosticReports/}), which includes an Emoji record and a (partial) NewWords record.}
\label{fig:report_file}
\end{figure}

\begin{figure}[h]
\includegraphics[scale=0.6]{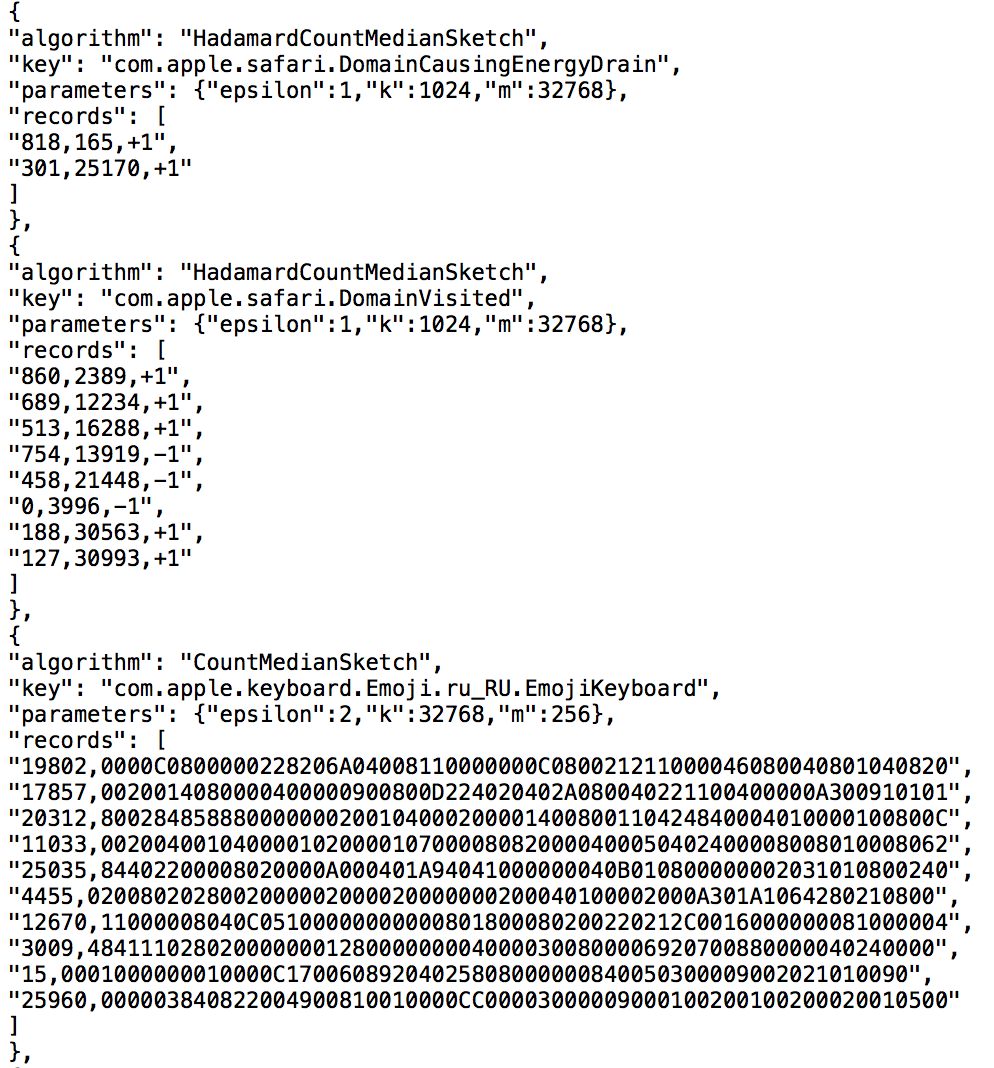}
\centering
\caption{An example report file from iOS11.}
\label{fig:report_fileiOS11}
\end{figure}

\begin{figure}[h]
\includegraphics[scale=0.6]{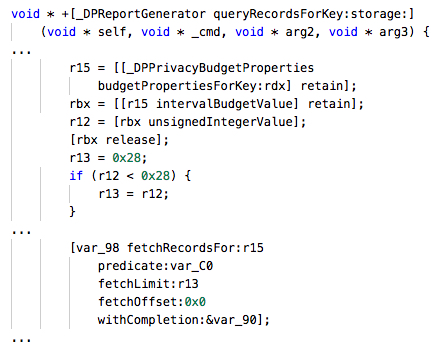}
\centering
\caption{Code snippet for fetch records to submit (0x28 = 40).}
\label{fig:min40}
\end{figure}

\begin{figure*}[h]
\includegraphics[scale=0.4]{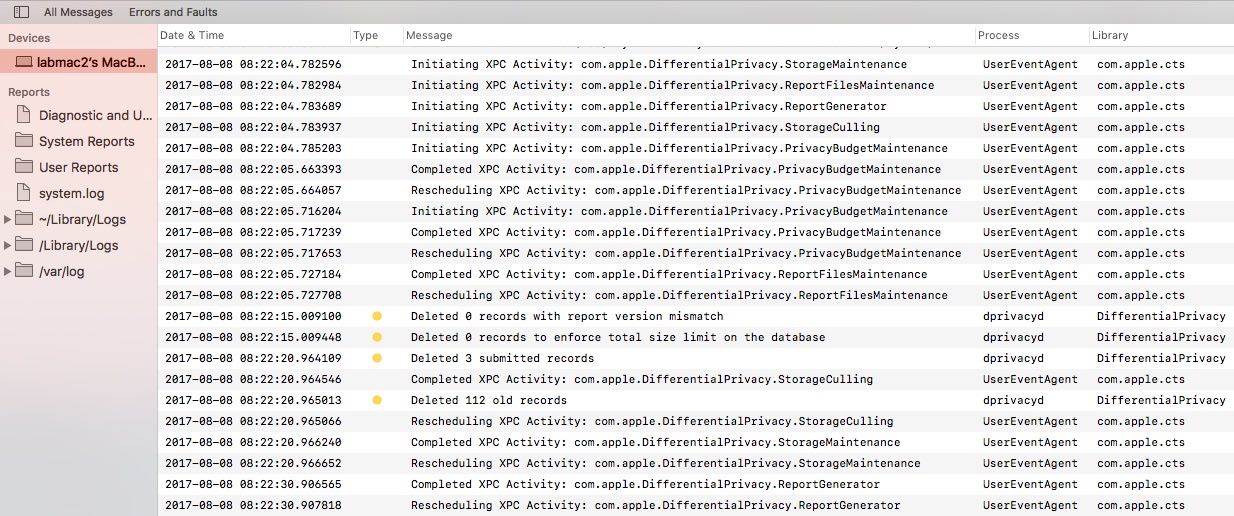}
\centering
\caption{A screenshot of Console.app, which includes output from 5 DP periodic tasks.}
\label{fig:new_console}
\vspace{-1em}
\end{figure*}


%


\end{document}